\documentclass[final,conference]{IEEEtran}
\newtheorem{theorem}{Theorem}
\newtheorem{dfi}{Definition}

\ifCLASSINFOpdf
\usepackage[pdftex]{graphicx}
\graphicspath{{../pdf/}{../jpeg/}}
\DeclareGraphicsExtensions{.pdf,.jpeg,.png}
\else
\usepackage[dvips]{graphicx}
 \graphicspath{{../eps/}}
 \DeclareGraphicsExtensions{.eps}
\fi
\usepackage[cmex10]{amsmath}
\newcommand{\mc}{\mathcal{C}}
\newcommand{\mb}{\mathcal{B}}
\newcommand{\rt}{\tilde{R}}
\newcommand{\mtx}{\mathcal{T_X}}
\newcommand{\mty}{\mathcal{T_Y}}
\newcommand{\mtz}{\mathcal{T_Z}}

\usepackage{mdwmath}
\usepackage{mdwtab}
\IEEEoverridecommandlockouts

\begin{document}
%
\title{Empirical Coordination in a Triangular Multiterminal Network
}
\author{\IEEEauthorblockN{Ali Bereyhi, Mohsen Bahrami, Mahtab Mirmohseni and Mohammad Reza Aref\thanks{This work was partially supported by Iranian NSF under contract no. $88114/46-2010$.}}
\IEEEauthorblockA{Information Systems and Security Lab (ISSL),\\
Sharif University of Technology, Tehran, Iran,\\
Email: \{bereyhi, bahramy\}@ee.sharif.edu, mirmohseni@ee.sharif.ir, aref@sharif.edu }
}


%


\maketitle

\begin{abstract}
In this paper, we investigate the problem of the empirical coordination in a triangular multiterminal network. A triangular multiterminal network consists of three terminals where two terminals observe two external i.i.d correlated sequences. The third terminal wishes to generate a sequence with desired empirical joint distribution. For this problem, we derive inner and outer bounds on the empirical coordination capacity region. It is shown that the capacity region of the degraded source network and the inner and outer bounds on the capacity region of the cascade multiterminal network can be directly obtained from  our inner and outer bounds. For a cipher system, we establish key distribution over a network with a reliable terminal, using the results of the empirical coordination. As another example, the problem of rate distortion in the triangular multiterminal network is investigated in which a distributed doubly symmetric binary source is available.
\end{abstract}

\section{Introduction}
Reconstruction of a source by means of limited resources is one of the primary purposes of communication. In \cite{1}, Shannon discussed the problem of lossless source coding where a source is intended to be transmitted over a rate limited noiseless channel and showed that the minimum required rate for source description is the entropy of the source. The description rate for distributed sources can be reduced if there is correlation between the sources. Slepian and Wolf established the optimal rate region of the lossless distributed source coding \cite{3}. In the problem of lossless source coding, the source sequences can be reconstructed at receivers without any distotion. The problem of realizing sequences with a specified distance from the source sequences is introduced by Shannon \cite{2}, where the rate distortion function is defined as a deterministic function of the distance and the source distribution. The rate distortion problem for two correlated sources was discussed by Berger and Tung \cite{4} and \cite{5}. In their model, two separate encoders intend to transmit two correlated sources over noiseless channels and a receiver tries to reconstruct the sources subject to corresponding distortions. The rate distortion problem in a cascade network was first studied by Yamamato \cite{6}. In such networks, there are three terminals-- a transmitter, a relay terminal and a receiver-- which are connected by two noiseless links in a cascade setting. Furthermore, the transmitter has access to an i.i.d source. For this model, the rate distortion capacity is derived where the relay terminal and the transmitter intend to reconstruct two distorted sequences of the source sequence. Permuter and Weissman established the rate distortion capacity region of the cascade and triangular networks where side information is available at the transmitter and the relay terminal \cite{7}. Chia \emph{et al.} extended the Permuter's model to a network in which a degraded side information is also available at the receiver \cite{8}. 

In the problem of rate distortion, the reconstructed sequences may have different statistics. In some cases, it is required to realize certain joint statistics between the reconstructed sequences and the source sequences. Cuff \emph{et al.} considered the coordination problem to achieve certain joint statistics between terminals in a network \cite{9}. Based on the definition of the statistics, there are two concepts of coordination referred to as empirical and strong coordination. In the empirical coordination problems, the terminals upon observing correlated sources, wish to generate sequences with desired empirical joint distribution. Generating sequences with a certain induced distribution in multiple terminal networks, where some terminals have access to correlated sources, is classified under the strong coordination problems. Cuff studied the empirical coordination in a cascade network where the transmitter and the relay terminal observe two correlated sources. The receiver utilizes the received message from the relay terminal to generate a sequence with given empirical joint distribution \cite{10}.

Many applications can be modeled as a coordination problem. For instance, consider a network in which multiple description sources try to make a terminal acts as a new source. This new source may need to act in some joint behavior with other sources to fit this network into another network through a certain bottleneck. As another application, assume that we intend to generate pseudo-random sequences with specific empirical distribution in a network. In these applications, we have to satisfy a terminal to generate a sequence with a desired distribution. One of the interesting applications of the empirical coordination is key distribution in cipher systems. Consider a cipher system with an encryptor and $n$ decryptors which are distributed over a network. In order to establish a secure connection over the network, the encryptor enciphers a plain text by means of a random key sequence. The encryptor intends to distribute the key sequence to the decryptors over the network using rate-limited secure channels. In addition $m$ reliable terminals are available in the network. The reliable terminals have access to some sequences, correlated with the key sequence. In the information theory context, we develop the problem of key distribution using the empirical coordination. We consider a reliable terminal which helps an encryptor to distribute a key sequence to a decryptor.

In this paper, we investigate a noiseless triangular network where two terminals, which have access to correlated sources, stimulate the third terminal to construct a sequence with desired empirical distribution as illustrated in Fig.~1. In some cases our model reduces to the cascade multiterminal and degraded source networks introduced in [9]. The cascade multiterminal network can be deduced by eliminating the direct link $\mathcal{C}_3$. Considering $Y$ as a deterministic function of $X$ yields the degraded source model. For the Triangular Multiterminal Network (TMN), inner and outer bounds on the empirical coordination capacity region are derived. The inner bound is established using two coding schemes. In each coding scheme, the Wyner-Ziv \cite{11} and superposition coding \cite{12} are utilized. The results are used to implement key distribution in a cipher system where a reliable terminal observes a sequence correlated with the key sequence. As another example, we discuss the problem of rate distortion in the TMN in which a doubly symmetric binary source is available.

The rest of the paper is organized as follows: In Section II, the problem definition is given. In Section III, we provide our main results and the intuitions behind them. In Section IV, we present the examples. Finally, proof of theorems are illustrated in Section V.
\section{Problem Definition}{
Throughout the paper, we denote a discrete random variable with an upper case letter (e.g., $X$) and its realization by the lower case letter (e.g., $x$). We denote the probability density function of $X$ over $\mathcal{X}$ with $p(x)$ and the conditional probability density function of $Y$ given $X$ by $p(y|x)$. We also use $X^n$ to indicate vector $(X_{1},X_{2}, \ldots ,X_{n})$.

A TMN consists of three terminals which are connected by three rate-limited noiseless channels, as Fig.~1 illustrates. $\mathcal{T_X}$ and $\mathcal{T_Y}$ have access to i.i.d sources $X^n$ and $Y^n$, respectively. The sources $X^n$ and $Y^n$ are correlated according to probability distribution $p(x,y)$. $\mathcal{T_X}$ can communicate over two noiseless channels $\mathcal{C}_1$ and $\mathcal{C}_3$ which are limited by rates $R_1$ and $R_3$. In addition, a noiseless channel $\mathcal{C}_2$ provides one way communication from $\mathcal{T_Y}$ to $\mathcal{T_Z}$ with limited rate $R_2$. $\mathcal{T_X}$ upon observing $x^n$ transmits the messages $m_1=m_1(x^n)$ and $m_3=m_3(x^n)$ over $\mathcal{C}_1$ and $\mathcal{C}_3$ to $\mathcal{T_Y}$ and $\mathcal{T_Z}$, respectively. $\mathcal{T_Y}$ after receiving $m_1$ and observing $y^n$ transmits $m_2=m_2(y^n , m_1(x^n))$ to $\mathcal{T_Z}$ over $\mathcal{C}_2$. By this scheme, $\mathcal{T_Y}$ acts as a relay with side information. $\mathcal{T_Z}$ generates a sequence $z^n$ as a deterministic function of received messages $m_2$ and $m_3$, i. e., $z^n=z^n(m_2(y^n,m_1(x^n)), m_3(x^n))$.

\begin{dfi}[Empirical Distribution]
Consider three deterministic sequences $x^n$, $y^n$ and $z^n$. The empirical distribution of $x^n$, $y^n$ and $z^n$ is defined as
\begin{align}
&\hat{p}_{x^n,y^n,z^n}(x,y,z)=\frac{\sum_{i=1}^n I\{(x_i,y_i,z_i)=(x,y,z)\}}{n},
\end{align}
where $I\{.\}$ is the indicator function.
\end{dfi}

\begin{dfi}[Total Variation]
Consider two probability distributions $p(x)$ and $q(x)$. The total variation distance is specified by
\begin{align}
& {\parallel p(x)-q(x) \parallel}_1=\sum_{i=1}^{\mid \mathcal{X} \mid} \mid p(x_i) - q(x_i) \mid.
\end{align}
\end{dfi}
\begin{figure}
\begin{center}
\includegraphics[width=3in,keepaspectratio]{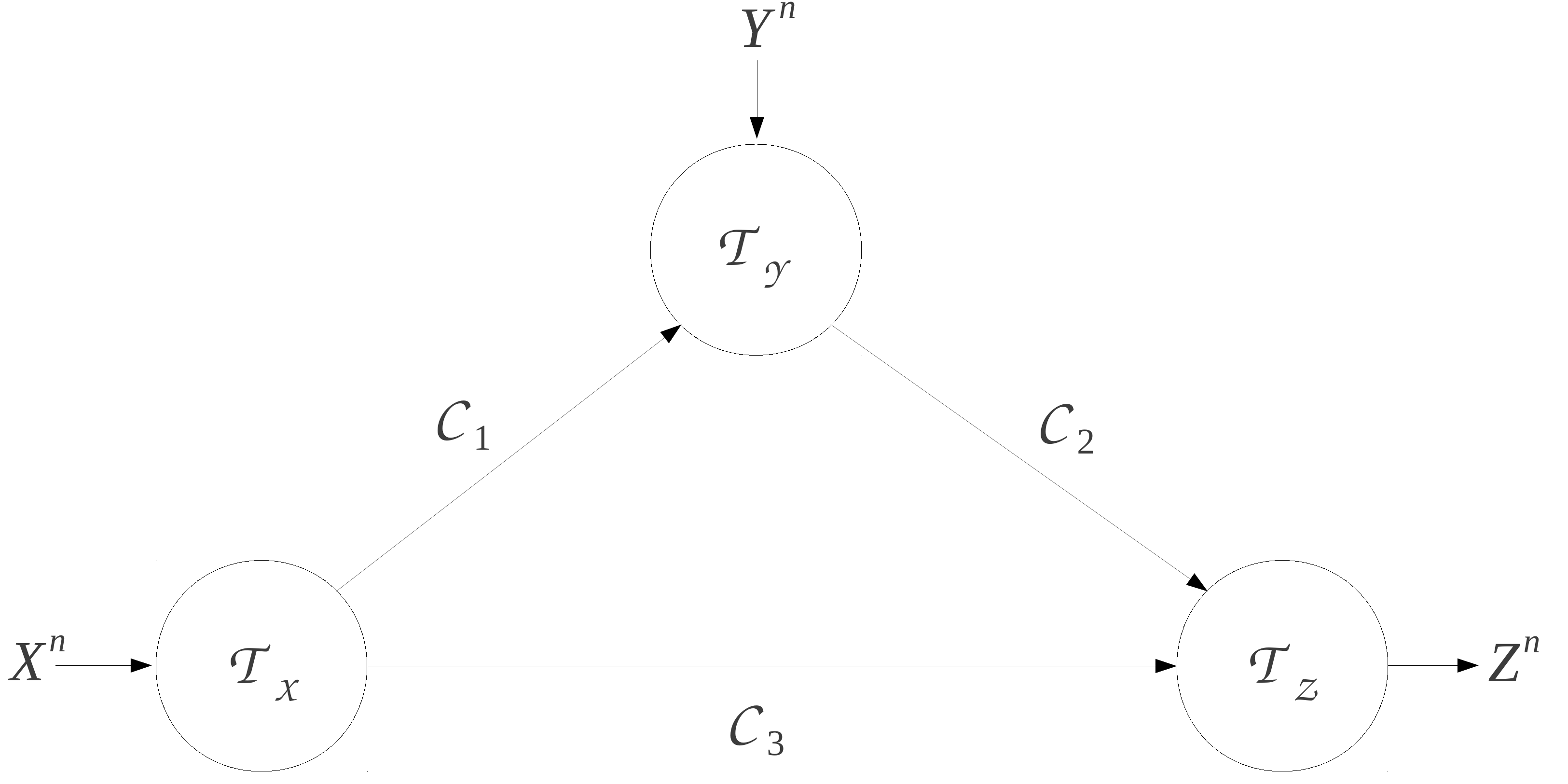}
\end{center}
\caption{The triangular multiterminal network}
\end{figure}
\begin{dfi}[Coordination Code]
A coordination code $(n, 2^{nR_1}, 2^{nR_2}, 2^{nR_3},f,r,g)$ for the TMN consists of an enc- oder {\small{$f: \mathcal{X}^n \to \{ 1,2, \ldots ,2^{nR_1} \} \times \{ 1,2, \ldots ,2^{nR_3} \}$}}, a recoding function {\small{$r: \{ 1,2, \ldots ,2^{nR_1} \} \times \mathcal{Y}^n \to \{ 1,2, \ldots ,2^{nR_2} \}$}} and a decoding function {\small{$g: \{ 1,2, \ldots ,2^{nR_2} \} \times \{ 1,2, \ldots ,2^{nR_3} \} \to \mathcal{Z}^n$}}.
\end{dfi}

\emph{Definition 4 (Empirical Coordination Achievability):} A joint probability distribution $p(x,y)p(z|x,y)$ is~said to be achievable empirically with the rate triple $(R_1,R_2,R_3)$, if there exists a sequence of coordination code $(n, 2^{nR_1}, 2^{nR_2}, 2^{nR_3},f,r,g)$, such that  ${\parallel \hat{p}_{X^n,Y^n,Z^n}(x,y,z)-p(x,y)p(z|x,y) \parallel}_1 \to 0$ in probability. The rate triple $(R_1,R_2,R_3)$ is called achievable rate triple.

\emph{Definition 5 (Empirical Coordination Capacity Region):} The set of all achievable rate triples $(R_1,R_2,R_3)$ is defined as empirical coordination capacity region.
}
\section{Main Results}{
In this section, we give the inner and outer bounds on the empirical coordination capacity region of the TMN in Theorem 1 and Theorem 2, respectively.

\begin{theorem}[Inner Bound]
For the desired joint distribution $p(x,y)p(z|x,y)$, the rate triple $(R_1,R_2,R_3)$ is achievable empirically in the TMN if $(R_1,R_2,R_3) \in \mathcal{R}$ where
\begin{align}
\mathcal{R}=ConvexHull(\mathcal{R}_1 \cup \mathcal{R}_2) \label{7}
\end{align}
and $\mathcal{R}_1$ and $\mathcal{R}_2$ are defined as follows:
\begin{align}
\mathcal{R}_1=\{ (&R_1,R_2,R_3) \ \ni \ R_1 \geq I(X;U,V|Y), \nonumber \\
&R_2 \geq I(X;U)+I(V,Y;Z|U)-I(W;Z|U), \nonumber \\
&R_3 \geq I(X;W|U) \} \label{3}
\end{align}
for some input distributions $p(x,y,z,u,v,w)=p(x,y)p(u|x)\\p(v|u,x)p(w|u,x)p(z|y,u,v,w)$,
and
\begin{align}
\mathcal{R}_2=\{ (&R_1,R_2,R_3) \ \ni \ R_1 \geq I(X;U,V|Y), \nonumber \\
&R_2 \geq I(X;U)+I(V,Y;W|U), \nonumber \\
&R_3 \geq I(X;Z|U)-I(W;Z|U) \} \label{4}
\end{align}
for some input distributions $p(x,y,z,u,v,w)=p(x,y)p(u|x)\\p(v|u,x)p(w|u,y)p(z|x,u,w)$.
\end{theorem}
\begin{IEEEproof}[Outline of the proof]
For the achievability, we utilize two encoding and decoding schemes. In each scheme, the Wyner-Ziv and superposition coding are used. In order to achieve $\mathcal{R}_1$, $\mtx$ generates i.i.d sequences $U^n$ and $V^n$, then randomly partitions them. $\mtx$ upon observing $X^n$ finds jointly typical $(U^n,V^n)$ with $X^n$ and transmits bin indices of $U^n$ and $V^n$ over $\mc_1$. Similarly, $\mtx$ generates sequences $W^n$ jointly typical with $U^n$ and sends index of a sequence, which is jointly typical with $X^n$, over $\mc_3$. $\mty$, after receiving the indices, first relays the index of $U^n$ to $\mtz$. Then, $\mty$ generates sequences $Z^n$ jointly typical with $U^n$ and partitions them using random binning. $\mty$ chooses a sequence $Z^n$ typical with $V^n$ and transmits the bin index of $Z^n$ over $\mathcal{C}_2$. $\mtz$ finds sequence $Z^n$ in the bin by means of received $W^n$. In this scheme, $W^n$ roles as side information. $\mathcal{R}_2$ can be obtained when, instead of $\mtx$, $\mty$ provides the side information. In this scheme, $\mtx$ utilizes $\mc_1$ like the first scheme and $\mty$ relays the index of $U^n$, however, the bin index of $Z^n$ which is typical with $X^n$ and $U^n$ is sent over $\mc_3$. In addition, $\mty$ finds $W^n$ typical with $(Y^n,V^n,U^n)$ and transmits its index over $\mc_2$. Similar to the first scheme, $\mtz$ chooses $Z^n$. The inner bound is deduced by convexity of empirical coordination capacity region. Detailed proof is provided in Section~\ref{sec:seca}.
\end{IEEEproof}

\emph{Remark 1}: As $X$ and $Y$ are correlated sources, variable $V$ is used by $\mty$ for compressing. The variables $W$ and $U$ are used by $\mtz$ for reconstruction. 

\emph{Remark 2:} In Theorem~1, by setting $W=\emptyset$ in $\mathcal{R}_1$ and $W=X$ in $\mathcal{R}_2$, the inner bound reduces to the inner bound on the empirical coordination capacity region of the cascade multiterminal network studied by Cuff \emph{et al.} [9].
}

\begin{theorem}[Outer Bound]
In order to achieve joint distribution $p(x,y)p(z|x,y)$ empirically in the TMN, the rate triple $(R_1,R_2,R_3)$ must satisfy
\begin{align}
&R_1 \geq I(X;U,V|Y) \nonumber \\
&R_2 \geq I(X,Y;U) \nonumber \\
&R_3 \geq I(X;W|U)
\end{align}
for some input distributions $p(x,y,z,u,v,w)$.
\end{theorem}

\emph{Proof}: See Section~\ref{sec:secb}. 

\emph{Remark 3:} By setting $U=Z$, $W=\emptyset$ and considering the input distribution as $p(x,y,z,v)=p(x,y)p(v|x)p(z|y,v)$ in Theorem~2, the outer bound reduces to the outer bound on the empirical coordination capacity region of the cascade multiterminal network investigated by Cuff \emph{et al.} [9].

\emph{Remark 4:} By assuming $Y=f(X)$, i.e., $Y$ is a deterministic function of $X$, and setting $V=W=\emptyset$ in Theorem~1 and $V=\emptyset$, $W=Z$ in Theorem~2, the region reduces to the empirical coordination capacity region of the degraded source model discussed by Cuff \emph{et al.} [9].
\section{Examples}
Different problems can be modeled as the problem of empirical coordination. In the following, we discuss some examples of such problems in the TMN.

Consider a TMN where $\mtx$ and $\mty$ observe i.i.d correlated sequences $X^n$ and $Y^n$, respectively. Let $(X,Y)$ be a Doubly Symmetric Binary Source (DSBS($a$)), i.e., $\Pr(X=0,Y=0)=\Pr(X=1,Y=1)= \frac{1}{2} a$ and $\Pr(X=0,Y=1)=\Pr(X=1,Y=0)= \frac{1}{2} (1-a)$, $a \in [0,\frac{1}{2}]$. For this network, we investigate two examples-- key distribution in a cipher system and the problem of rate distortion in the TMN.
\subsection{Key Distribution in a Cipher System}
Suppose a cipher system where there are an encryptor, a reliable terminal and a decryptor. In this system, the encryptor enciphers a plain text with the key sequence $X^n$. The reliable terminal, which observes the sequence $Y^n$, has the ability to communicate over a secure noiseless channel $\mc_2$ with the decryptor. The encryptor intends to share the key sequence with the decryptor by sending the required information over a noiseless channel $\mc_3$. Also, the encryptor has access to a noiseless channel $\mc_1$ to communicate with the reliable terminal. The encryptor has some limits on secure communication, therefore, desires to save its output sum-rate as much as possible. We model this system by the TMN and obtain an inner bound on $(R_1,R_2,R_3)$, where $R_i$ is the transmission rate that is sent over $\mc_i$, for $i=1,2,3$.

Let $p(u|x)$ be a Binary Symmetric Channel (BSC($\alpha$)) and consider $V=W=\emptyset$ in Theorem~1, the set $\mathcal{R}_2$ is deduced as,
\begin{align}
R_1 &\geq I(X;U,V|Y)=I(X;U|Y)=H(U|Y)-H(U|X,Y) \nonumber \\
&=H_b(a \ast \alpha)-H_b(\alpha) \nonumber \\
R_2 &\geq I(X;U)+I(V,Y;W|U)=I(X;U)=1-H_b(\alpha) \nonumber \\
R_3 &\geq I(X;Z|U)-I(W;Z|U)=H(X|U)=H_b(\alpha) \nonumber
\end{align}
and we obtain:
\begin{align}
&R_1 \geq H_b(a \ast \alpha)-H_b(\alpha) \nonumber \\
&R_2 \geq 1-H_b(\alpha) \nonumber \\
&R_3 \geq H_b(\alpha), \nonumber
\end{align}
for some {\small{$\alpha \in [0,\frac{1}{2}]$}}, where {\small{$H_b(x)=-x \log x - (1-x) \log (1-x)$}} and {\small{$x \ast y = x(1-y)+y(1-x)$}}.

The above expressions state that if we set $a=\frac{1}{2}$, the reliable terminal acts as a relay. In fact, in this condition $X$ and $Y$ are independent variables and the reliable terminal can not help the encryptor to save its output sum-rate. On the other hand if we choose $a=0$, $R_1$ can be equal to $0$. In This condition $Y=X \oplus 1$ and the reliable terminal can generate the key sequence by complementing $Y^n$. Fig.~2 illustrates the variation of the encryptor's output sum-rate with respect to $a$. From the Fig.~2, it is clear that the output sum-rate increases by increasing the value of $a$.
\begin{figure}
\includegraphics[width=3.7in,keepaspectratio]{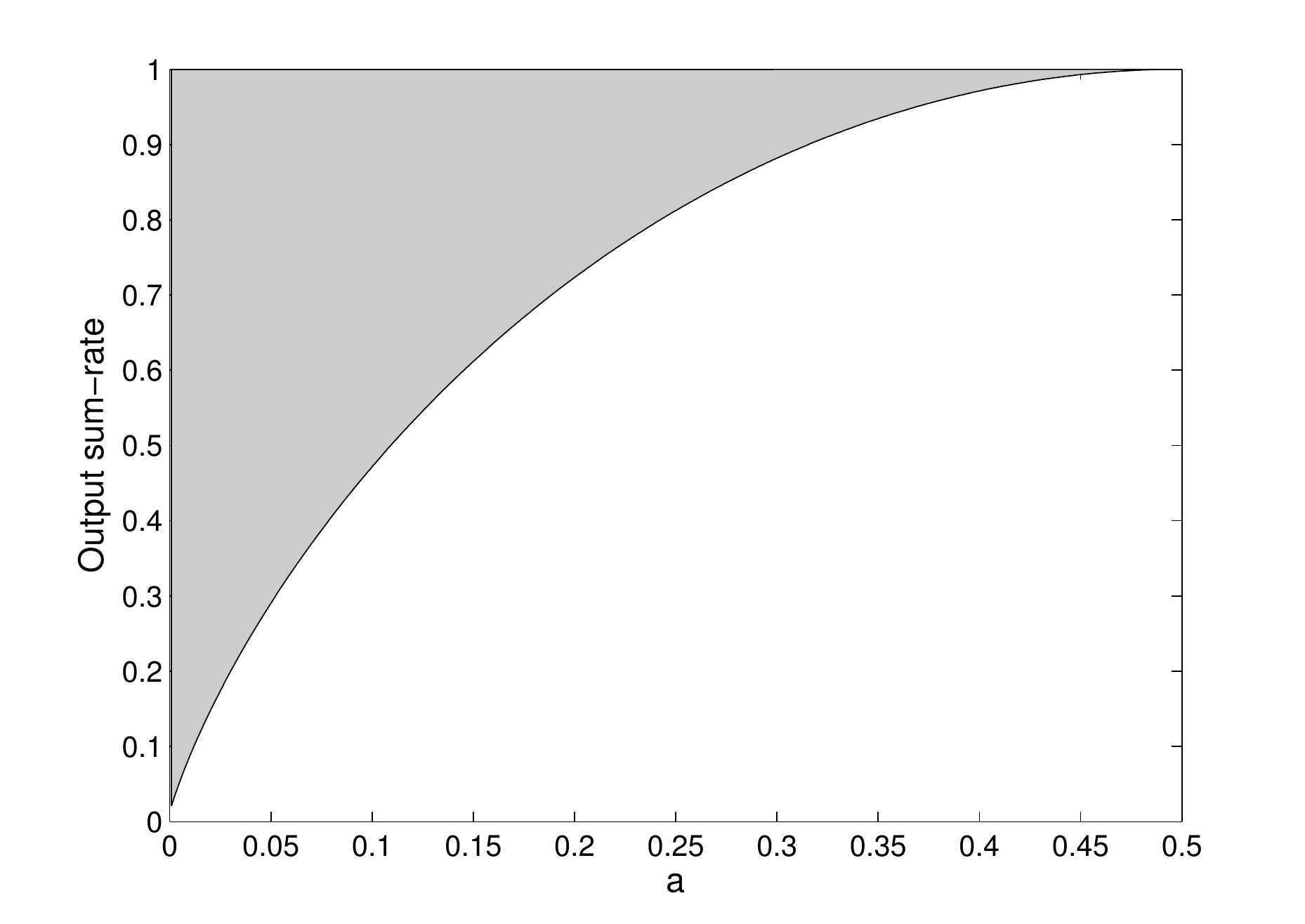}
\label{5}
\caption{The output sum-rate variation respect to $a$. For $a=0$ sum-rate can be equal to $0$, in this condition the reliable terminal can individually generate key sequence by complementing $Y^n$. For $a=\frac{1}{2}$ the reliable terminal has no information about the key sequence and dose not help.}
\end{figure}
\subsection{The Problem of Rate Distortion in The TMN}
For the proposed TMN, we investigate the problem of rate distortion where $\mtz$ intends to reconstruct $X^n$ with maximum distortion $D$ according to the distortion function $d(x,\hat{x})=x \oplus \hat{x}$. In this network, $Y^n$ roles as side information.

Using Theorem 1, let $p(u|x)$ be a BSC($\alpha$) and $p(\hat{x}|u,y)=p(\hat{x}|y)$  be a BSC($d$) in $\mathcal{R}_1$. By setting $V=W=\emptyset$, we obtain
\begin{align}
R_1 &\geq I(X;U,V|Y)=I(X;U|Y)=H_b(a \ast \alpha)-H_b(\alpha), \nonumber \\
R_2 &\geq I(X;U)+I(Y,V;Z|U)-I(W;Z|U)=I(X;U) \nonumber \\
&+I(Y;\hat{X}|U) =2-H_b(\alpha)-H_b(d), \nonumber \\
R_3 &\geq I(X;W|U)=0, \nonumber
\end{align}subject to the constraint $a \ast d \leq D$.

For $\mathcal{R}_2$, let $p(u|x)$ be a BSC($\alpha$) and $p(\hat{x}|u,x)=p(\hat{x}|x)$ be a BSC($d$). By setting $V=W=\emptyset$, we obtain
\begin{align}
R_1 &\geq I(X;U,V|Y)=I(X;U|Y)=H_b(a \ast \alpha)-H_b(\alpha), \nonumber \\
R_2 &\geq I(X;U)+I(Y,V;W|U)=1-H_b(\alpha), \nonumber \\
R_3 &\geq I(X;Z|U)-I(W;Z|U)=I(X;\hat{X}|U) \nonumber \\
&=H(\hat{X}|U)-H(\hat{X}|X,U)=H_b(\alpha \ast d)-H_b(d), \nonumber
\end{align}subject to the constraint $d\leq D$.

By defining $\mathcal{D}_1$ and $\mathcal{D}_2$ as
\begin{align}
\mathcal{D}_1&= \{ (R_1,R_2,R_3) \ni R_1 \geq H_b(a \ast \alpha)-H_b(\alpha), \nonumber \\
&R_2 \geq 2-H_b(\alpha)-H_b(d), \ R_3 \geq 0 \ \ \forall d: a \ast d \leq D \}, \nonumber\\
\mathcal{D}_2&= \{ (R_1,R_2,R_3) \ni R_1 \geq H_b(a \ast \alpha)-H_b(\alpha), \nonumber \\
&R_2 \geq 1-H_b(\alpha), \ R_3 \geq H_b(\alpha \ast d)-H_b(d) \ \forall d:d \leq D \}, \nonumber
\end{align}
for some $\alpha \in [0,\frac{1}{2}]$, we have
\begin{align}
&\mathcal{D}_1 \cup \mathcal{D}_2 \subseteq R(D).
\end{align}

For this problem, consider a case that the variables $X$ and $Y$ are independent, i.e., $a=\frac{1}{2}$. In this case $\mathcal{D}_1=\emptyset$ if $D \neq \frac{1}{2}$ and $\mathcal{D}_1 \cup \mathcal{D}_2=\mathcal{D}_2$. In fact when $X$ and $Y$ are independent, $\mty$ roles as a relay. At the other side, if $X^n$ can be generated completely from $Y^n$, i.e., $a=0$, $\mc_1$ is not needed. In this case, by setting $\alpha =\frac{1}{2}$ in $\mathcal{D}_1$, the source can be described with $R_2 \geq 1-H_b(D)$ and $R_1=R_3=0$. Also, by setting $\alpha =\frac{1}{2}$ in $\mathcal{D}_2$, the source can be described with $R_3 \geq 1-H_b(D)$ and $R_1=R_2=0$. In fact in this case both $\mtx$ and $\mty$ have the source sequence.
\section{Proofs}{
In this section, we present the achievability and converse proofs for the described model. We employ the Wyner-Ziv and the superposition coding for the achievability proof. In order to achieve the inner bound, two different coding schemes are established. The time mixing trick is used in the proof of the outer bound [9]. Before we state the proofs, we illustrate lemma~1 [13, p. 37]. This lemma lets us define the typical set by means of the total variation distance.

\emph{Lemma~1~(\cite{13})} Define the set $\mathcal{T}^{(n)}_\epsilon(X)$ for distribution $p(x)$ as {\small{$\mathcal{T}^{(n)}_\epsilon(X)= \{ x^n \ni {\parallel \hat{p}_{x^n}(x)-p(x) \parallel}_1 \leq \epsilon \}$}}. The set {\small{$\mathcal{T}^{(n)}_\epsilon(X)$}} is bounded as {\small{$\mathcal{A}^{(n)}_\epsilon \subseteq \mathcal{T}^{(n)}_\epsilon(X) \subseteq \mathcal{A}^{(n)}_{\epsilon \mid \mathcal{X} \mid}$}}, where {\small{$\mathcal{A}^{(n)}_\epsilon$}} denotes the strongly typical set for distribution {\small{$p(x)$}}. 

The proof can be directly deduced from the definitions of $\mathcal{T}^{(n)}_\epsilon(X)$ and strongly typical set $\mathcal{A}^{(n)}_\epsilon$.

This lemma indicates that for the finite set $\mathcal{X}$, the total variation between the distribution $p(x)$ and the empirical distribution of a typical sequence is small enough.
\subsection{Proof of Theorem~1} \label{sec:seca}
{
Fix a joint distribution $p(u,v,w,z)=p(u)p(v|u)p(w|u)$ $p(z|u)$.

\emph{Codebook Generation}: Generate $2^{n\tilde{R}_U}$ sequences $U^n(m_u)$, ${m_u \in [1:2^{n\tilde{R}_U})}$ each according to $\prod_{i=1}^np(u_i)$ and partition them into $2^{nR_U}$ bins. In each bin there are $2^{n(\tilde{R}_U - R_U)}$ sequences $U^n$ in average. For each sequence $U^n(m_u)$ randomly and conditionally independently generate $2^{n\tilde{R}_V}$, $2^{n\tilde{R}_W}$ and $2^{n\tilde{R}_Z}$ sequences $V^n(m_v)$, ${m_v \in [1:2^{n\tilde{R}_V})}$, $W^n(m_w)$, ${m_w \in [1:2^{n\tilde{R}_W})}$ and $Z^n(m_z), m_z \in {[1:2^{n\tilde{R}_Z})}$ according to distributions $\prod_{i=1}^np(v_i|u_i)$, $\prod_{i=1}^np(w_i|u_i)$ and $\prod_{i=1}^np(z_i|u_i)$, respectively. Then, randomly partition the sequences $V^n$ and $Z^n$ into $2^{nR_V}$ and $2^{nR_Z}$ bins, therefore, in each bin there are $2^{n(\tilde{R}_V - R_V)}$ and $2^{n(\tilde{R}_Z - R_Z)}$ sequences, respectively.

The codebook containing all sequences $U^n$ is shown by $\mathcal{C}^U$ and the corresponding bins with $\mathcal{B}^U(b_u), \ b_u \in [1:2^{nR_U})$. For each $U^n(m_u)$ we show the corresponding sub-codebooks consisting all sequences $V^n$, $W^n$ and $Z^n$ with $\mathcal{C}^V(m_u)$, $\mc^W(m_u)$ and $\mc^Z(m_u)$, respectively. Each bin of $\mc^V(m_u)$ is represented by $\mb^V(b_v), \ b_v \in [1:2^{nR_V})$. Also, we present each bin of $\mc^Z(m_u)$ by $\mb^Z(b_z), \ b_z \in [1:2^{nR_Z})$.
\subsection*{The First Scheme}
\emph{Encoding at $\mtx$}: Upon observing the source sequence $x^n$, $\mtx$ chooses a sequence $u^n(m_u) \in \mc^U$ such that $(x^n,u^n(m_u))$ are jointly typical. $\mtx$ chooses a sequence $v^n(m_v) \in \mc^V(m_u)$ such that $(x^n,u^n(m_u),v^n(m_v))$ are jointly typical. In addition, a sequence $w^n(m_w) \in \mc^W(m_u)$ is chosen such that $(x^n,u^n(m_u),w^n(m_w))$ are jointly typical. Then, the bin indices $b_u$ and $b_v$ are transmitted over the channel $\mc_1$ where $u^n(m_u) \in \mb^U(b_u)$ and $v^n(m_v) \in \mb^V(b_v)$, respectively. $\mtx$ transmits $m_w$ over the channel $\mc_3$. By the covering lemma \cite{15}, this can be done with an arbitrarily small probability of error as $n \to \infty$ if $\rt_U \geq I(U;X)$, $\tilde{R}_V \geq I(V;X|U)$ and $\tilde{R}_W \geq I(W;X|U)$.

\emph{Decoding at $\mty$}: $\mty$ reconstructs ${u^n(\hat{m}_u) \in \mb^U(b_u)}$ by using the observed sequence $y^n$ and the received bin index $b_u$ such that $(u^n(\hat{m}_u), y^n)$ are jointly typical. $\mty$ estimates $ v^n(\hat{m}_v) \in \mb^V(b_v)$ by using $y^n$ and the received bin index $b_u$ such that $(u^n(\hat{m}_u),v^n(\hat{m}_v),y^n)$ are jointly typical. By the packing and mutual packing lemma \cite{15}, the probability of error tends to zero as $n \to \infty$ if $\rt_U-R_U \leq I(U;Y)$, $\tilde{R}_V - R_V \leq I(V;Y|U)$ and $(\rt_U-R_U) +(\tilde{R}_V-R_V) \leq I(U,V;Y)$.

\emph{Encoding at $\mty$}: After decoding $u^n(\hat{m}_u)\ \text{and}\ v^n(\hat{m}_v)$, $\mty$ chooses $z^n(m_z)\in \mc^Z(\hat{m}_u)$ by using $y^n$ and the decoded sequences such that $(z^n(m_z),u^n(\hat{m}_u),v^n(\hat{m}_v),y^n)$ are jointly typical. $\mty$ transmits $ \hat{m}_u$ and the bin index $b_z$ over the channel $\mc_2$ where $z^n(m_z) \in \mb^Z(b_z)$. By the covering lemma, this can be done with an arbitrarily small probability of error as $n \to \infty$ if $\rt_Z \geq I(Z;V,Y|U)$.

\emph{Decoding at $\mtz$}: After receiving $m_w, \ \hat{m}_u$ and $b_z$, $\mtz$ reconstructs $w^n(m_w) \in \mc^W(\hat{m}_u)$ and $z^n(\hat{m}_z) \in \mb^Z(b_z)$ such that $(z^n(\hat{m}_z),w^n(m_w),u^n(\hat{m}_u))$ are jointly typical. By the packing lemma, the probability of error tends to zero as $n \to \infty$ if $\rt_Z - R_Z \leq I(Z;W|U)$.

Consequently, the rate triple $(R_1,R_2,R_3)$ can be written as $(R_1,R_2,R_3)=(R_U+R_V,\rt_U+R_Z,\rt_W)$. Using the Fourier-Motzkin elimination and considering the above equations, we get the expressions in \eqref{3}.
}
\subsection*{The Second Scheme}{
\emph{Encoding at $\mtx$}: $\mtx$ chooses $u^n(m_u)$ and $v^n(m_v)$, similar to the first scheme. A sequence $z^n(m_z)\in \mc^Z(m_u)$ is chosen such that $(z^n(m_z) , u^n(m_u), x^n)$ are jointly typical. $\mtx$ transmits the bin indices $b_u$ and $b_v$ over $\mc_1$ and the bin index $b_z$ over $\mc_3$ such that ${u^n(m_u) \in \mb^U(b_u)}$, ${v^n(m_v) \in \mb^V(b_v)}$ and ${z^n(m_z) \in \mb^Z(b_z)}$. By the covering lemma, this can be done with an arbitrarily small probability of error as $n \to \infty$ if $\rt_U \geq I(U;X)$, $\tilde{R}_V \geq I(V;X|U)$ and $\tilde{R}_Z \geq I(Z;X|U)$.

\emph{Decoding at $\mty$}: $\mty$ reconstructs $ u^n(\hat{m}_u)$ and $v^n(\hat{m}_v)$ similar to the first scheme.

\emph{Encoding at $\mty$}: After decoding $u^n(\hat{m}_u)$ and $v^n(\hat{m}_v)$, $\mty$ chooses ${w^n(m_w) \in \mc^W(\hat{m}_u)}$ using the observed sequence $y^n$ and the decoded sequences such that $(w^n(m_w),u^n(\hat{m}_u), v^n(\hat{m}_v),y^n)$ are jointly typical. $\mty$ transmits $ \hat{m}_u \ \text{and} \ m_w$ over $\mc_2$. By the covering lemma, this can be done with an arbitrarily small probability of error as $n \to \infty$ if $\rt_W \geq I(W;V,Y|U)$.

\emph{Decoding at $\mtz$}: After receiving $m_w$, $\hat{m}_u$ and $b_z$, $\mtz$ reconstructs $w^n(m_w) \in \mc^W(\hat{m}_u)$ and $z^n(\hat{m}_z) \in \mb^Z(b_z)$, similar to the first scheme. By the packing lemma, the probability of error tends to zero as $n \to \infty$ if $\rt_Z - R_Z \leq I(Z;W|U)$.

Consequently, the rate triple $(R_1,R_2,R_3)$ can be written as $(R_1,R_2,R_3)=(R_U+R_V,\rt_U+\rt_W,R_Z)$. Using the Fourier-Motzkin elimination and considering the above equations, we get the expressions in \eqref{4}. The convexity of the empirical coordination capacity region \cite{13} deduces \eqref{7}.
}
}
\subsection{Proof of Theorem~2} \label{sec:secb}
{
In order to prove the outer bound, we utilize the time mixing trick. The random time variable $Q$ is uniformly distributed over $[1:n]$. First, consider $R_1$:
\begin{align}
&nR_1 \geq H(M_1) \geq H(M_1|Y^n)\stackrel{(a)}{=} H(M_1|Y^n)\nonumber \\
&+ H(M_2|M_1,Y^n) \stackrel{(b)}{=} H(M_1,M_2|Y^n)-H(M_1,M_2|Y^n,X^n) \nonumber \\
&=I(M_1,M_2;X^n|Y^n)=H(X^n|Y^n)-H(X^n|Y^n,M_1,M_2) \nonumber \\
&\stackrel{(c)}{=} \sum_{i=1}^nH(X_i|Y_i)-H(X_i|Y^n,X^{i-1},M_1,M_2) \nonumber \\
&\geq \sum_{i=1}^nH(X_i|Y_i)-H(X_i|Y^i,X^{i-1},M_1,M_2) \nonumber \\
&= \sum_{i=1}^n I(X_i;M_1,M_2,X^{i-1},Y^{i-1}|Y_i) \nonumber \\
&\stackrel{(d)}{=} \sum_{i=1}^n I(X_i;U_i,V_i|Y_i)=nI(X_Q;U_Q,V_Q|Y_Q,Q) \nonumber \\
&\stackrel{(e)}{=} nI(X_Q;U_Q,V_Q,Q|Y_Q)\geq nI(X_Q;U_Q,V_Q|Y_Q) \nonumber
\end{align}\\ where $(a)$ follows from the fact that $M_2$ is a deterministic function of $M_1$ and $Y^n$. $(b)$ is due to the fact that $M_1$ and $M_2$ are deterministic functions of $X^n$ and $Y^n$. $(c)$ is directly obtained from i.i.d distribution of $(X^n,Y^n)$. By defining $U_i = (M_2,X^{i-1},Y^{i-1})$ and $V_i = (M_1,X^{i-1},Y^{i-1})$, $(d)$ can be deduced. Finally, $(e)$ comes from the time mixing properties. Now, consider $R_2$:
\begin{align}
&nR_2 \geq H(M_2) \stackrel{(a)}{=} H(M_2)-H(M_2|X^n,Y^n) \nonumber \\
&=I(M_2;X^n,Y^n)+\sum_{i=1}^n I(X_i,Y_i;M_2|X^{i-1},Y^{i-1}) \nonumber \\
&= \sum_{i=1}^n I(X_i,Y_i;M_2|X^{i-1},Y^{i-1}) + I(X_i,Y_i;X^{i-1},Y^{i-1}) \nonumber \\
&= \sum_{i=1}^n I(X_i,Y_i;M_2,X^{i-1},Y^{i-1})= \sum_{i=1}^n I(X_i,Y_i;U_i) \nonumber \\
&=nI(X_Q,Y_Q;U_Q|Q)= nI(X_Q,Y_Q;U_Q,Q) \nonumber \\
&\geq nI(X_Q,Y_Q;U_Q) \nonumber
\end{align}\\ where $(a)$ follows from the fact that $M_2$ is a deterministic functions of $X^n$ and $Y^n$. Finally, for $R_3$ we have:
\begin{align}
nR_3 &\geq H(M_3) \geq H(M_3|M_2)\stackrel{(a)}{=} H(M_3|M_2) \nonumber  \\
&- H(M_3|M_2,X^n,Y^n)=I(M_3;X^n,Y^n|M_2) \nonumber 
\end{align}
\begin{align}
&= \sum_{i=1}^n H(X_i,Y_i|X^{i-1},Y^{i-1},M_2) \nonumber \\
&-\sum_{i=1}^n H(X_i,Y_i|X^{i-1},Y^{i-1},M_2,M_3)\stackrel{(b)}{=} \sum_{i=1}^nH(X_i,Y_i|U_i) \nonumber \\
&- \sum_{i=1}^n H(X_i,Y_i|U_i,W_i)= \sum_{i=1}^n I(X_i,Y_i;W_i|U_i) \nonumber \\
&\geq \sum_{i=1}^n I(X_i;W_i|U_i)=nI(X_Q;W_Q|U_Q,Q) \nonumber \\
&\geq nI(X_Q;W_Q|U_Q) \nonumber
\end{align}where $(a)$ follows from the fact that $M_3$ is a deterministic functions of $X^n$. By defining $W_i = (M_3,X^{i-1})$, $(b)$ can be deduced.
\section{Conclusion}{
We investigated the empirical coordination problem in a triangular network where the transmitter and the relay terminal observe two correlated sources. For this problem, inner and outer bounds on the empirical coordination capacity region were derived. In the achievability proof, two different coding schemes were used to provide two regions. The convex hull of these regions achieved the inner bound of capacity region.
}

\end{document}